\documentclass[final,1p,times]{elsarticle}

\usepackage{color}
\usepackage[colorlinks = true,
            linkcolor = blue,
            urlcolor  = blue,
            citecolor = blue,
            anchorcolor = blue]{hyperref}
            
\usepackage{pdflscape}
\usepackage{amssymb}
\usepackage[tight-spacing=true]{siunitx}
\usepackage{booktabs}
\usepackage{tabularx}
\usepackage{dcolumn}  
\DeclareSIUnit\wn{\cm\tothe{-1}}
\DeclareSIUnit\bar{bar}
\DeclareSIUnit{\atmosphere}{atm}
\DeclareSIUnit{\nothing}{\relax}

\newcolumntype{d}[1]{D{.}{.}{#1}} 

\biboptions{sort&compress}

\usepackage{amsthm}

\journal{Icarus}
\begin{document}

\begin{frontmatter}



\title{\textcolor{blue}{Self and N$_2$ collisional broadening of far-infrared methane lines at low-temperature with application to Titan.}}


\author[ICB]{C. Richard\corref{cor1}}\ead{Cyril.Richard@u-bourgogne.fr}
\author[ICB]{V. Boudon}
\author[SOLEIL,LISA]{L. Manceron}
\author[ULB]{J. Vander Auwera}
\author[LESIA]{S. Vinatier}
\author[LESIA]{B. Bézard}
\author[LESIA,GEN]{M. Houelle}

\address[ICB]{Laboratoire Interdisciplinaire Carnot de Bourgogne, UMR 6303 CNRS - Universit\'e Bourgogne Franche-Comt\'e, 9 Av. A. Savary, BP 47870, F-21078 Dijon Cedex, France}
\address[SOLEIL]{AILES, Synchrotron SOLEIL, L’Orme des Merisiers Saint-Aubin, F-91192 Gif-sur-Yvette, France}
\address[LISA]{LISA, Université Paris Cité and Univ Paris Est Creteil, CNRS, LISA, F-75013 Paris, France}
\address[ULB]{SQUARES, C.P. 160/09, Université Libre de Bruxelles, B-1050 Brussels, Belgium}
\address[LESIA]{LESIA, Observatoire de Paris, Université PSL, Sorbonne Université, Université Paris Cité, CNRS, 5 place Jules Janssen, 92195 Meudon, France}
\address[GEN]{Observatoire de Genève, Université de Genève, Chemin Pegasi 51, 1290 Sauverny, Switzerland}

\cortext[cor1]{\textcolor{blue}{Corresponding author}}

\begin{abstract}
We report the measurement of broadening coefficients of pure rotational lines of methane at different pressure and temperature conditions. A total of 27 far-infrared spectra were recorded at the AILES beamline of the SOLEIL synchrotron at room-temperature, \SI{200}{\kelvin} and \SI{120}{\kelvin}, in a range of 10 to \SI{800}{\milli \bar}. Self and N$_2$ broadening coefficients and temperature dependence exponents of methane pure rotational lines have been measured in the 73--\SI{136}{\wn} spectral range using multi-spectrum non-linear least squares fitting of Voigt profiles. These coefficients were used to model spectra of Titan that were compared to a selection of equatorial Cassini/CIRS spectra, showing a good agreement for a stratospheric methane mole fraction of (1.17 $\pm$ 0.08)\%.
\end{abstract}


\end{frontmatter}


\section{Introduction}
Methane is an important molecule for the study of planetary atmospheres. It is present in many planets and bodies of the Solar System, but it is also found, at high temperature, in the atmosphere of several exoplanets. The study of methane lines at low temperature, as in this study, is of great interest to probe the thick atmosphere of Titan~\cite{lellouch2010sounding} but also of ice giants. It has for example been used to measure the spatial distribution of methane in the stratosphere of Titan from the Cassini Composite InfraRed Spectrometer (CIRS) spectra~\cite{lellouch2014distribution}, showing here the importance of close-up observations. Although since 2018 no more spacecraft are into orbit around the Saturnian system, Dragonfly is a planned NASA mission~\cite{barnes2021science} which is expected to land on Titan in 2034, bringing to its surface a robotic rotorcraft. Moreover, interest in ice giants has been renewed recently with a proposed mission that plans to send a spacecraft into orbit around Uranus together with a descent probe, scheduled for launch in 2031 or 2032~\cite{national2022origins}. This planet is known to have methane clouds at a temperature of about \SI{100}{\kelvin}~\cite{sromovsky2008methane} showing again the interest to study the broadening of the methane lines at such temperatures. Methane has also been observed in Neptune’s stratosphere~\cite{orton2007evidence} and has been extensively analyzed with the PACS instrument onboard the \emph{Herschel} telescope~\cite{lellouch2010first}.

However, in the atmospheric environments, methane is generally mixed with other gases, like nitrogen in Titan's case. It is therefore essential to also study the broadening of methane lines in a medium diluted with nitrogen in order to take into account all collision broadening effects.

Section~\ref{sec:exp} details the experimental setup and the conditions used to record the spectra at the synchrotron facility SOLEIL. The measurements of nitrogen-broadening coefficients and temperature dependence exponents, and details on the fitting procedure are described in Section~\ref{sec:analysis} and Section~\ref{sec:rot_dep}. Section~\ref{sec:discussion} is a discussion of our results and compares them with values found in HITRAN and other references. Finally, we applied these results to an analysis of methane rotational lines observed on Titan in Section~\ref{sec:titan} and present our conclusions in Section~\ref{sec:conclusion}.

\section{Experimental details}\label{sec:exp}

To estimate the temperature dependence of the broadening coefficients of the very weak pure rotation lines of methane~\cite{Boudon10}, a wide range of temperature and pressure conditions combined with a long optical path length were required. Absorption spectra were therefore recorded on the AILES Beamline at the SOLEIL synchrotron facility (Saint-Aubin, France), with the synchrotron light source coupled to a Bruker IFS 125HR Fourier transform spectrometer~\cite{RBRp08,Brubach2010}, maintained at \SI{4E-5}{\milli\bar}, and the AILES cryogenic long path cell~\cite{KWLp13}. A total of nineteen high-resolution spectra were recorded at low-temperature with an optical path length of 93.14(1)~\si{\metre}. For the new spectra the iris was set to \SI{2.5}{\milli\metre}, while for the older spectra, the iris was open to it's maximum value of \SI{12.5}{\milli\metre}. In any case, the effective size of the source at the spectrometer entrance was smaller than the maximum size required for the highest resolution at the useful wavenumbers. The unapodized spectra were obtained in the 40--\SI{300}{\wn} region at resolutions varied from 0.05 to \SI{0.001}{\wn} (defined as 0.9/maximum optical path difference, see Table~\ref{Tab:exp_conditions}), with a \SI{3.8}{\centi\meter/\second} scanner velocity, a \SI{6}{\micro\meter} Si/Mylar beamsplitter and a \SI{4}{\kelvin}-cooled Si composite bolometer with a \SI{1.5}{\milli\second} rise time and a cold \SI{300}{\wn} low-pass filter. The recorded spectra result from the co-addition of various numbers of interferograms (see Table~\ref{Tab:exp_conditions}). They were divided by low resolution (\SI{0.05}{\wn}) empty cell spectra taken at the same temperature. The resulting transmittance spectra were zero-filled, corrected for channelling effects and calibrated using well-known water rovibrational lines~\cite{gordon2022hitran2020} with a standard deviation of about \SI{3E-4}{\wn}. 

Two series of low-temperature measurements were carried out specifically for this study (Table~\ref{Tab:exp_conditions}), while spectra at room temperature were recorded previously by \citet{sanzharov2012self} and are not reported in this study. For all these spectra, the synchrotron radiation from the SOLEIL facility (\SI{500}{\milli\ampere} in the most stable operation mode with 316 equally filled electron bunches, \SI{430}{\milli\ampere} for spectra 4 and 5) was used as a source. Spectra 4 and 5, with the highest pressures of pure methane, were recorded in a first series at an earlier date. For these measurements, the objective mirrors were slightly warmed by specific heaters to keep mirror actuators from jarring, and this resulted in a larger pressure gradient which is responsible for relatively large temperature error bars. In a second measurement series at a later date (spectra 1-3 and 6-19), this effect could be drastically reduced and the uncertainties from the temperature gradient are reduced by a factor of two. Uncertainty on methane or nitrogen broadening gas pressures is always smaller than 0.5\%.

\renewcommand{\arraystretch}{1.25}
\begin{table}[ht!]
\caption{Experimental conditions for the low-temperature CH$_4$ spectra recorded in this study.}\label{Tab:exp_conditions}
\vspace{5pt}
\resizebox{1.0\hsize}{!}{\begin{tabular}{SSSSSS}
\hline\hline
 \multicolumn{1}{c}{Spectrum} & \multicolumn{1}{c}{CH$_4$ Pressure$^\text{a}$ / \si{\milli\bar}} & \multicolumn{1}{c}{Temperature$^\text{a}$ / \si{\kelvin}} & \multicolumn{1}{c}{\# Averaged scans} & \multicolumn{1}{c}{Resolution / \si{\wn}} & \multicolumn{1}{c}{Pressure N$_2$$^\text{a}$ / \si{\milli\bar}} \\
 \hline
1 & 10.67 (5) & 120 (4) & 336 & 0.001 & 0 \\
2 & 20.4 (1) & 120 (4) & 320 & 0.002 & 0 \\
3 & 41.7 (2) & 125 (4) & 256 & 0.004 & 0 \\
4 & 50.0 (3) & 129 (10) & 578 & 0.005 & 0 \\
5 & 100.1 (5) & 117 (10) & 320 & 0.01 & 0 \\
6 & 21.3 (1) & 124 (4) & 1000 & 0.005 & 79.5 (4) \\
7 & 41.7 (2) & 125 (4) & 320 & 0.008 & 159.0 (8) \\
8 & 80.0 (4) & 120 (4) & 1060 & 0.02 & 320 (2)\\
9 & 79.5 (4) & 125 (4) & 320 & 0.01 & 80 (4) \\
10 & 160.0 (8) & 130 (4) & 840 & 0.05 & 638 (3) \\
11 & 10.12 (5) & 199 (2) & 320 & 0.0012 & 0 \\
12 & 20.5 (1) & 199 (2) & 504 & 0.0025 & 0 \\
13 & 42.6 (2) & 199 (2) & 432 & 0.005 & 0 \\
14 & 83.5 (4) & 199 (2) & 432 & 0.01 & 0 \\
15 & 20.4 (1) & 199 (2) & 936 & 0.005 & 80.5 (4) \\
16 & 40.85 (2) & 199 (2) & 432 & 0.01 & 161.0 (8)\\
17 & 80.7 (4) & 199 (2) & 576 & 0.005 & 80.3 (4) \\
18 & 80.1 (4) & 199 (2) & 480 & 0.02 & 320 (2)\\
19 & 160.2 (8) & 199 (2) & 1200 & 0.05 & 640 (3)\\
\hline\hline
\multicolumn{5}{l}{\footnotesize{$^\text{a}$ Estimated uncertainty in parentheses.}}
\end{tabular}}
\end{table}

\section{Retrieval of broadening coefficients and their temperature dependence exponents}\label{sec:analysis}

\subsection{Fitting procedure}\label{sec:fitting_procedure}

\begin{figure*}[ht]
  \centerline{\resizebox{1.0\hsize}{!}{\includegraphics{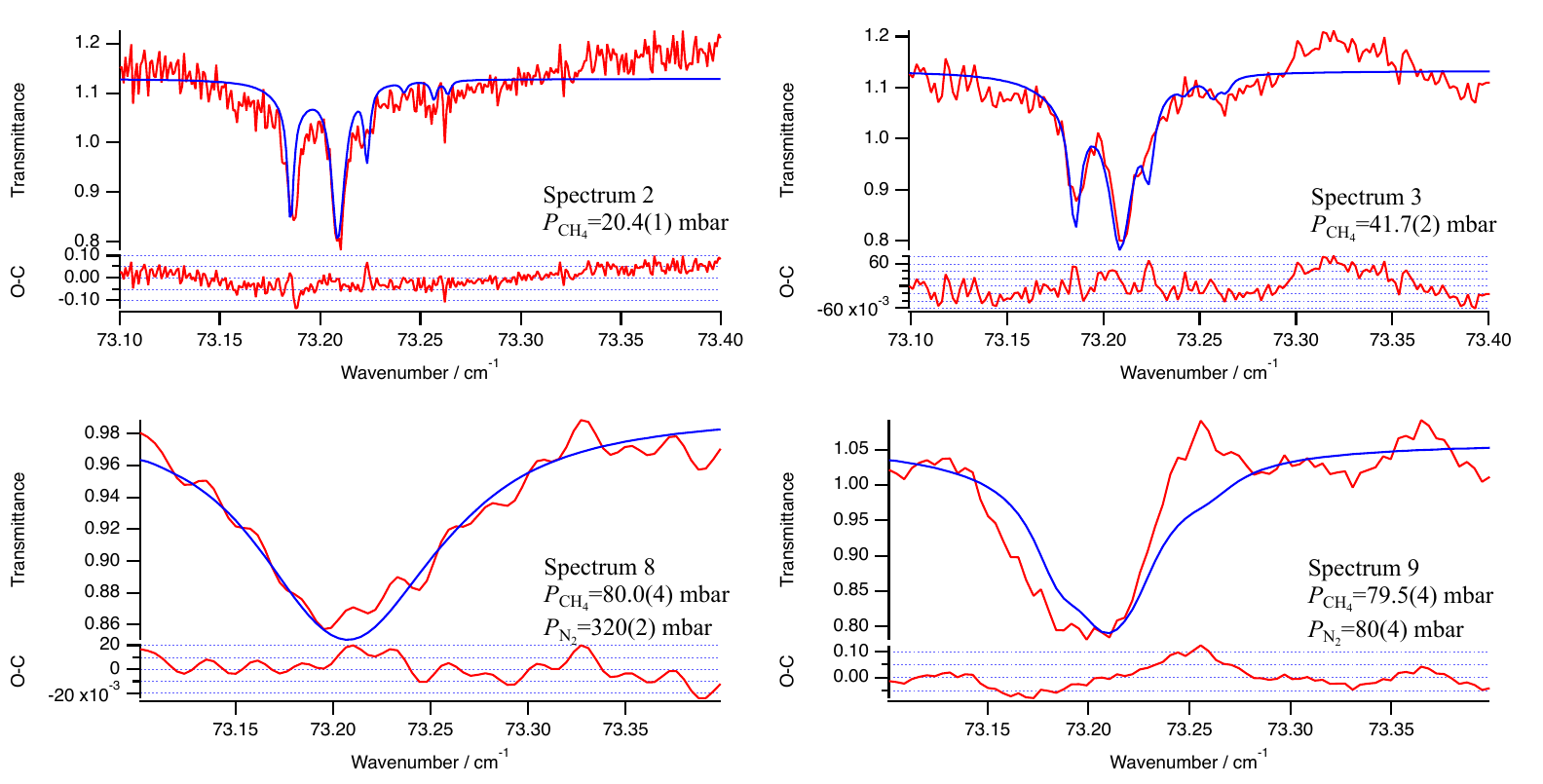}}}
\caption{Multispectrum non-linear fit of the \emph{R}(6) manifold: the 4 observed spectra, in red, (see Table~\ref{Tab:exp_conditions}) included in the fit, in blue, are overlaid with the corresponding best-fit calculated spectra and residuals (lower pannels).}
\label{fig:R_6_fitting}
\end{figure*}

\begin{figure*}[ht]
  \centerline{\resizebox{1.0\hsize}{!}{\includegraphics{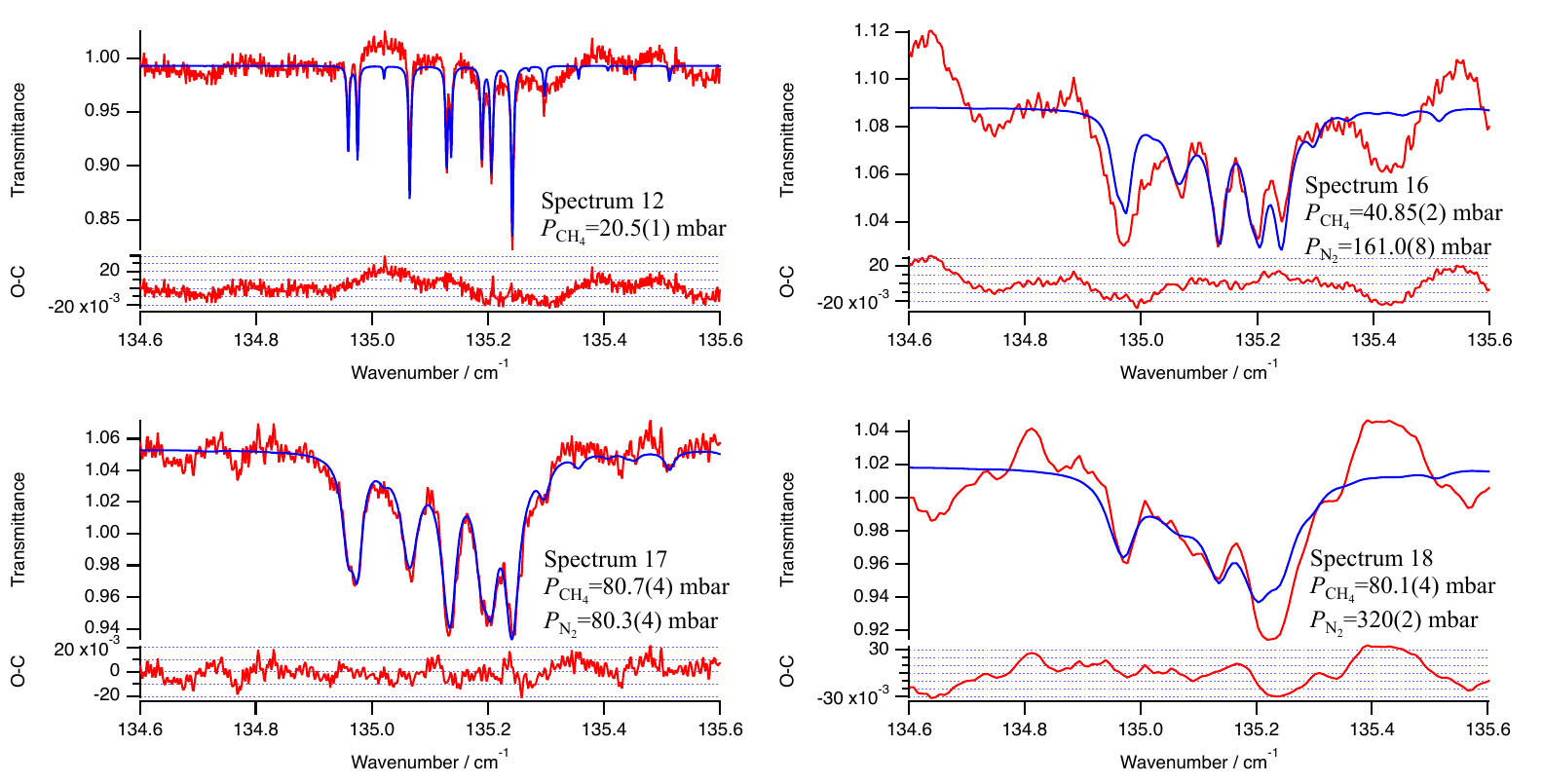}}}
\caption{Multispectrum non-linear fits of \emph{R}(12) lines (see Table~\ref{Tab:exp_conditions} for experimental conditions).}
\label{fig:R_12_fitting}
\end{figure*}

The CH$_4$-N$_2$ collisional half widths were measured using a multi-spectrum non-linear least squares fitting program~\cite{tudorie2012co2, daneshvar2014infrared}. The CH$_4$ manifolds were successively treated, all the spectra available for each of them being fitted simultaneously. The measurements typically involved the simultaneous adjustment of 17 to 27 observed spectra for each manifold. Each calculated spectrum was computed as the convolution of a monochromatic transmission spectrum with an instrument line shape function, which includes the effects of the finite maximum optical path difference and of the finite source aperture diameter of the interferometer. The background in each spectrum was represented by a polynomial expansion up to the second order (a constant or a linear function was however found to be sufficient in most cases), and the profile of the lines was modeled using a Voigt function with Gaussian width always held fixed to the value calculated for the Doppler broadening. Line intensities were fixed to the values calculated by \citet{boudon2010high} and listed in the HITRAN database~\cite{brown2013methane}. Pressure-induced line shift was not needed to fit the spectra to the noise level and line-mixing effects were neglected. The temperature dependent pressure induced widths of the methane lines were modeled according to the following expression:
\begin{equation}
\gamma_L = P_{tot} \, \left[ \gamma_{\mathrm{N_2}} \, \left( \frac{T_0}{T}\right)^{n_{\mathrm{N_2}}} (1-x) + \gamma_\mathrm{self} \, \left( \frac{T_0}{T}\right)^{n_{\mathrm{self}}} x \right]
\end{equation}
where $P_{tot}$ is the total sample pressure, $T_0 = \SI{296}{\kelvin}$, $T$ is the temperature (in \si{K}) and $x$ is the methane mole fraction (\textit{i.e.} the ratio of the methane to total pressure). $\gamma_{\mathrm{N_2}}$ and $\gamma_\mathrm{self}$ are the N$_2$ and self broadening coefficients at $T_0$, respectively. $n_{\mathrm{N_2}}$ and $n_{\mathrm{self}}$ are their respective unitless temperature dependence exponents. Examples of best fits are given in Fig.~\ref{fig:R_6_fitting} and Fig.~\ref{fig:R_12_fitting} for the \emph{R}(6) and \emph{R}(12) lines.

The parameters presented in this paper were retrieved from multispectrum non-linear fits. The fit was first performed on pure methane spectra data at room temperature and then integrating the low temperatures (spectra 1-5, 11-14 in Table~\ref{Tab:exp_conditions}). Finally, when the fit was satisfactory, the data with the nitrogen mixture were integrated (spectra 6-10, 15-19 in Table~\ref{Tab:exp_conditions}). All parameters remained free and were calculated simultaneously.

\begin{figure}[ht]
\centerline{\resizebox{1.0\hsize}{!}{\includegraphics{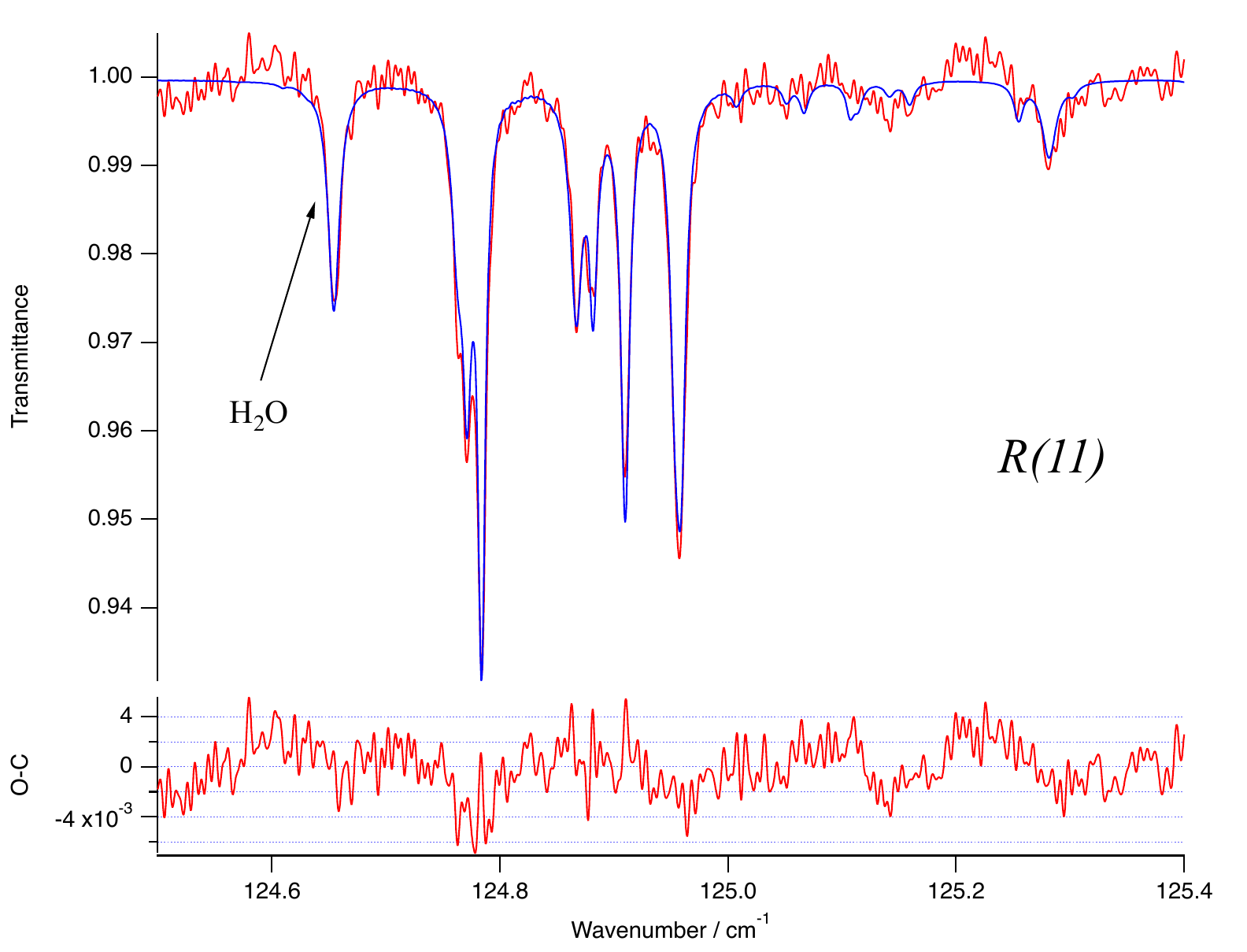}}}
\caption{Analysis of the \emph{R(11)} region of the far-infrared spectrum of $^{12}$CH$_4$, taken at room temperature with a mixture of nitrogen at \SI{133}{\milli \bar} (see Ref.~\cite{sanzharov2012self}). The observed spectrum is shown in the upper part, in red and the fitted spectrum in blue. The fit residuals are presented in the lower panel. A water vapor line, included in the fit, is also identified.}
\label{fig:R11_H2O}
\end{figure}

Water vapor lines overlap some of the CH$_4$ manifolds. For example, H$_2$O lines measured at \SI{73.2622}{\wn} and \SI{124.6536}{\wn} are very close to the \emph{R(6)} and \emph{R(11)} manifolds, respectively. Fig.~\ref{fig:R11_H2O} illustrates the latter situation. The positions and intensities of these water vapor lines were fixed to the values available in the HITRAN database \cite{gordon2022hitran2020}, while the H$_2$O mole fraction and an effective broadening parameter of these lines were fitted.

The N$_2$ and self broadening coefficients and their temperature dependence exponents measured in this work are reported in Table~\ref{tab:all_param}. They concern 45 lines belonging to the \emph{R(6)} to \emph{R(12)} manifolds of $^{12}$CH$_4$. For data that could not be fitted, HITRAN values were taken, and are given without uncertainty.

\subsection{Uncertainty estimation}

Estimating the accuracy of the measured broadening coefficients requires considering the uncertainties on the physical parameters, contributions from possible systematic errors as well as the uncertainties derived from the fits, taken as the standard deviation. The dominant contributions to the systematic errors ($\varepsilon_{\text{sys}}$) arise from the location of the full-scale photometric level, channeling, electronic and detector nonlinearities. An arbitrary value of 5\% has been retained for $\varepsilon_{\text{sys}}$, much greater than the 2\% value in a recent study~\cite{attafi2019self} given the lesser stability of the long path measurements at cryogenic temperatures. Upper limits of the overall relative uncertainties on the measured parameters $\varepsilon_{\mathrm{total}}^{rel}$ were calculated as the sum in quadrature (thus assuming they are uncorrelated) of the maximum relative uncertainties on the individual experimental parameters, {\em i.e.\/} $\varepsilon_{\text{si}}$ (sample purity, 0.1\%), $\varepsilon_{\text{T}}$ (temperature, ranging from 3.5 \% at \SI{120}{\kelvin} to 1\% at \SI{200}{\kelvin}, except for spectra 4 and 5 where uncertainties were estimated to 8\%), $\varepsilon_{\text{p}}$ (pressure, 0.5\%), $\varepsilon_{\text{pl}}$ (pathlength, 0.01\%), $\varepsilon_{\text{fit}}$ (standard deviation from the fit, given for each coefficient in Table~\ref{tab:all_param}). The estimated overall uncertainties $\varepsilon_{\mathrm{total}}$ listed in Table~\ref{tab:all_param} were then calculated as the product of $\varepsilon_{\mathrm{total}}^{rel}$ with the corresponding parameter. For the sake of simplicity, and because the difference would be relatively small, we have used the upper limit of $\varepsilon_{\text{T}}=3.5\%$ for all coefficients. 

\begin{figure*}[ht]
  \centerline{\resizebox{1.0\hsize}{!}{\includegraphics{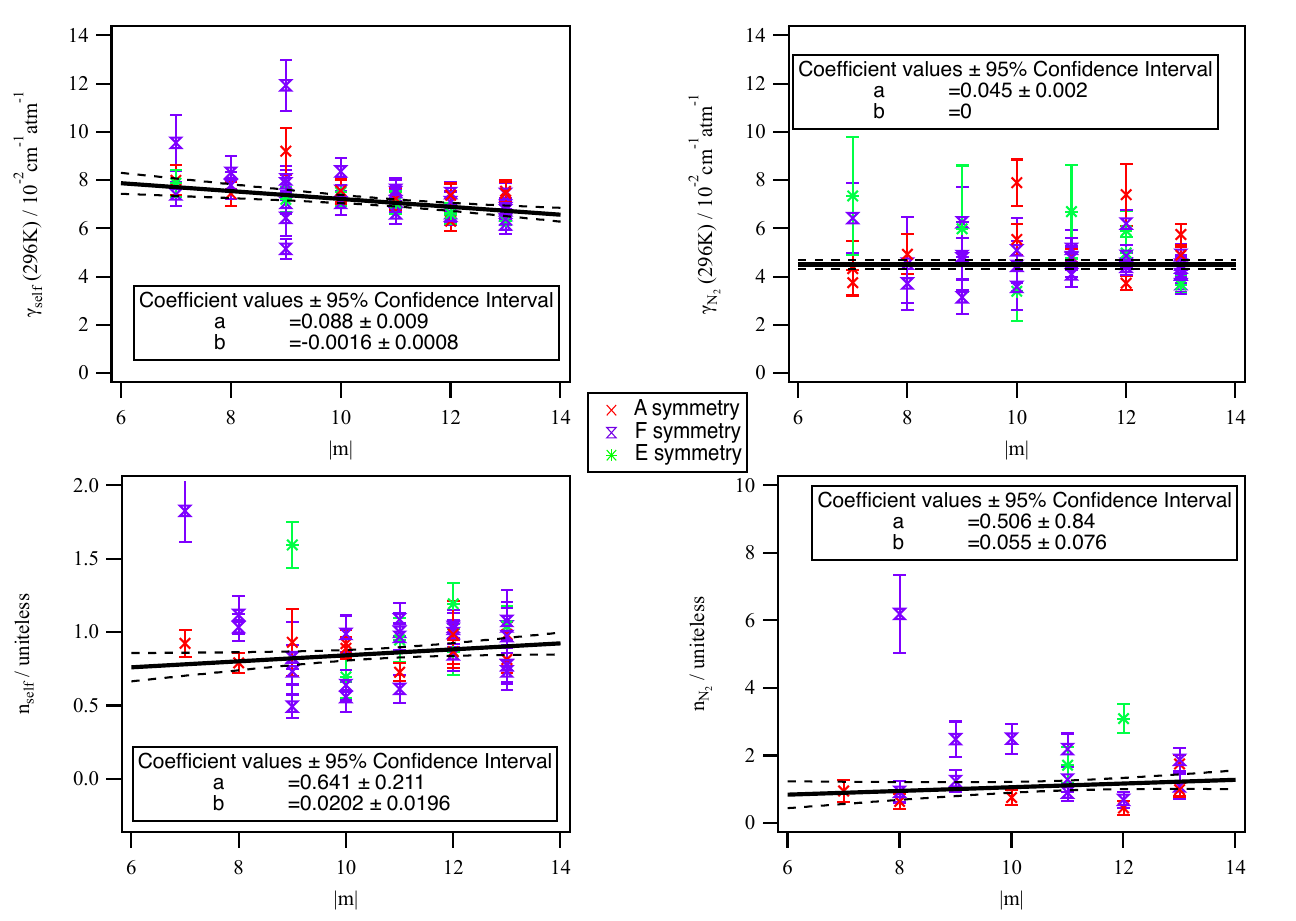}}}
\caption{Evolution with $|m|$ (where $m=J''+1$ for \emph{R}-branch lines) of self and N$_2$ broadening parameters and temperature dependence exponents. Different symbols and colors are used for the $A$-, $E$- and $F$-species transitions. Weighted fits were performed on each set of data points to obtain coefficients used in section~\ref{sec:titan}. One data point is excluded from the plot of $\gamma_{\mathrm{N_2}}$ at $|m|=7$ for a better visualization.}
\label{fig:sel_n2_br}
\end{figure*}

\begin{landscape}
\begin{table*}[ht]
 \caption{Measured N$_2$ and self broadening coefficients and temperature exponents for pure rotational lines of $^{12}$CH$_4$ in natural abundance. The line positions are expressed in \si{\wn}, while line intensities  are in \si{10\tothe{-25} \wn / (molecule.\cm\tothe{-2})} at \SI{296}{\kelvin}. For data that could not be fitted, HITRAN values were taken, and are given without uncertainty. The last six columns give the line assignments (using $''$ for the lower sates and $'$ for the upper states).}
 \label{tab:all_param}

~

\centering
\resizebox{\textwidth}{!}{\begin{tabular}{
	S
	S
        c
	S
	S
	S
	S
	S
	S
	S
	S
	S
	S
        S
	S
	S
	c
	S
	S
	c
	S
	}
\hline\hline
\multicolumn{1}{c}{\text{Position}} & \multicolumn{1}{c}{\text{Intensity}} & \multicolumn{1}{c}{\text{Temperature used (K)}} & \multicolumn{1}{c}{\text{$\gamma_{\text{N}_2}$}} & \multicolumn{1}{c}{$\varepsilon_{\text{fit}}$} & \multicolumn{1}{c}{$\varepsilon_{\text{total}}$} & \multicolumn{1}{c}{\text{$n_{\text{N}_2}$}} & \multicolumn{1}{c}{$\varepsilon_{\text{fit}}$} & \multicolumn{1}{c}{$\varepsilon_{\text{total}}$} & \multicolumn{1}{c}{\text{$\gamma_{\text{self}}$}} & \multicolumn{1}{c}{$\varepsilon_{\text{fit}}$} & \multicolumn{1}{c}{$\varepsilon_{\text{total}}$} & \multicolumn{1}{c}{\text{$n_{\text{self}}$}} & \multicolumn{1}{c}{$\varepsilon_{\text{fit}}$} & \multicolumn{1}{c}{$\varepsilon_{\text{total}}$} & \multicolumn{1}{c}{\text{$J'$}} & \multicolumn{1}{c}{\text{$C'$}} & \multicolumn{1}{c}{\text{$\alpha'$}} & \multicolumn{1}{c}{\text{$J''$}} & \multicolumn{1}{c}{\text{$C''$}} & \multicolumn{1}{c}{\text{$\alpha''$}}\\
\hline
  73.186198 &  0.9201 &  120/200/298 & 0.37950    &  0.15000  &  0.15179 &  0.700 &        &      & 0.09547  &   0.01000 & 0.01158  &  1.830 & 0.180 & 0.212 & 7 &  $F_2$ & 1 & 6 & $F_1$ & 1 \\
  73.208941 &  2.0010 & 120/200/298 & 0.04327    &  0.01100  &  0.01131 &  0.952 &  0.31  & 0.32  & 0.07992  &   0.00390 & 0.00626  &  0.924 & 0.070 & 0.090 & 7 &  $A_2$ & 1 & 6 & $A_1$ & 1 \\
  83.569132 &  1.3050 & 120/200/298 & 0.04534    &  0.01900  &  0.01920 &  6.420 &  1.10  & 1.16 &  0.08310  &   0.00470 & 0.00693  &  1.120 & 0.110 & 0.130 & 8 &  $F_1$ & 1 & 7 & $F_2$ & 2 \\
  83.576220 &  2.8620 & 120/200/298 & 0.04920    &  0.00770  &  0.00827 &  0.638 &  0.22  & 0.22 &  0.07446  &   0.00210 & 0.00502  &  0.792 & 0.048 & 0.068 & 8 &  $A_1$ & 1 & 7 & $A_2$ & 1 \\
  83.607176 &  1.8830 & 120/200/298 & 0.03696    &  0.00760  &  0.00803 &  0.926 &  0.31  & 0.32 &  0.07791  &   0.00290 & 0.00558  &  1.040 & 0.067 & 0.092 & 8 &  $F_2$ & 1 & 7 & $F_1$ & 2 \\
  93.915549 &  1.6450 & 120/200/298 & 0.06255    &  0.01400  &  0.01451 &  2.480 &  0.50  & 0.52 &  0.08024  &   0.00260 & 0.00557  &  0.732 & 0.076 & 0.088 & 9 &  $F_2$ & 1 & 8 & $F_1$ & 2 \\
  93.931069 &  2.0780 & 120/200/298 & 0.04757    &  0.00790  &  0.00842 &  1.240 &  0.32  & 0.33 &  0.07053  &   0.00190 & 0.00472  &  0.830 & 0.061 & 0.079 & 9 &  $F_1$ & 1 & 8 & $F_2$ & 1 \\
  93.977670 &  1.7380 & 120/200/298 & 0.05984    &  0.02600  &  0.02626 &  0.690 &        &      &  0.07178  &   0.00300 & 0.00534  &  1.600 & 0.120 & 0.155 & 9 &  $E$  & 1 & 8 & $E$  & 2 \\
  93.978997 &  2.1910 & 120/200/298 & 0.04847    &  0.01400  &  0.01431 &  0.690 &        &      &  0.07878  &   0.00260 & 0.00548  &  0.494 & 0.073 & 0.079 & 9 &  $F_1$ & 2 & 8 & $F_2$ & 2 \\
  94.028716 &  0.4316 & 120/200/298 & 0.05600    &           &          &  0.690 &        &      &  0.06572  &   0.00660 & 0.00760  &  0.827 & 0.240 & 0.245 & 9 &  $F_1$ & 3 & 8 & $F_2$ & 2 \\
  94.142465 &  0.5664 & 120/200/298 & 0.05300    &           &          &  0.690 &        &      &  0.09282  &   0.00800 & 0.00971  &  0.931 & 0.220 & 0.227 & 9 &  $A_2$ & 1 & 8 & $A_1$ & 1 \\
 104.224694 &  3.2380 & 120/200/298 & 0.07893    &  0.00830  &  0.00961 &  0.650 &        &      &  0.07531  &   0.00180 & 0.00495  &  0.922 & 0.076 & 0.095 &10 &  $A_2$ & 1 & 9 & $A_1$ & 1 \\
 104.247364 &  2.1690 & 120/200/298 & 0.05087    &  0.01300  &  0.01337 &  0.650 &        &      &  0.07059  &   0.00230 & 0.00490  &  0.989 & 0.110 & 0.126  &10 &  $F_2$ & 1 & 9 & $F_1$ & 2 \\
 104.252281 &  1.6420 & 120/200/298 & 0.03389    &  0.01200  &  0.01218 &  0.650 &        &      &  0.07497  &   0.00320 & 0.00560  &  0.695 & 0.140 & 0.146 &10 &  $E$  & 1 & 9 & $E$  & 1 \\
 104.315064 &  2.7660 & 120/200/298 & 0.03576    &  0.00940  &  0.00965 &  2.500 &  0.42  & 0.45 &  0.08364  &   0.00260 & 0.00574  &  0.641 & 0.092 & 0.100 &10 &  $F_2$ & 2 & 9 & $F_1$ & 3 \\
 104.319236 &  2.5500 & 120/200/298 & 0.04454    &  0.01000  &  0.01037 &  0.650 &        &      &  0.07566  &   0.00240 & 0.00522  &  0.560 & 0.093 & 0.099 &10 &  $F_1$ & 1 & 9 & $F_2$ & 2 \\
 104.350038 &  5.5390 & 120/200/298 & 0.05540    &  0.00560  &  0.00655 &  0.758 &  0.21  & 0.22 &  0.07563  &   0.00140 & 0.00484  &  0.890 & 0.050 & 0.074 &10 &  $A_1$ & 1 & 9 & $A_2$ & 1 \\
 114.523421 &  2.2260 & 120/200/298 & 0.05153    &  0.00710  &  0.00777 &  2.190 &  0.44  & 0.46 &  0.07127  &   0.00150 & 0.00462  &  1.000 & 0.110 & 0.126 &11 &  $F_2$ & 1 &10 & $F_1$ & 1 \\
 114.535302 &  2.5860 & 120/200/298 & 0.04096    &  0.00460  &  0.00524 &  1.290 &  0.36  & 0.37 & 
 0.06618  &   0.00120 & 0.00423  &  1.090 & 0.091 & 0.113 &11 &  $F_1$ & 1 &10 & $F_2$ & 2 \\
 114.614369 &  1.8140 & 120/200/298 & 0.06697    &  0.01900  &  0.01944 &  1.710 &  0.53  & 0.54 &  0.07056  &   0.00200 & 0.00476  &  0.948 & 0.140 & 0.152 &11 &  $E$  & 1 &10 & $E$  & 2 \\
 114.617121 &  2.8020 & 120/200/298 & 0.04809    &  0.00760  &  0.00815 &  0.640 &        &      &  0.07505  &   0.00150 & 0.00483  &  0.614 & 0.089 & 0.097 &11 &  $F_2$ & 2 &10 & $F_1$ & 2 \\
 114.639394 &  6.9310 & 120/200/298 & 0.04796    &  0.00170  &  0.00339 &  0.640 &        &      &  0.07237  &   0.00084 & 0.00451  &  0.728 & 0.040 & 0.060 &11 &  $A_2$ & 1 &10 & $A_1$ & 1 \\
 114.650565 &  2.3870 & 120/200/298 & 0.03745    &  0.00480  &  0.00532 &  0.700 &        &      &  0.08500  &           &          &        &       &       & 7 &  $A_2$ & 2 & 6 & $A_1$ & 1 \\
 114.671461 &  3.5380 & 120/200/298 & 0.04760    &  0.00390  &  0.00487 &  0.902 &  0.26  & 0.27 &  0.07598  &   0.00110 & 0.00478  &  0.970 & 0.071 & 0.093 &11 &  $F_1$ & 2 &10 & $F_2$ & 3 \\
 114.926380 &  1.4430 & 120/200/298 & 0.06424    &  0.01400  &  0.01454 &  0.700 &        &      &  0.07409  &   0.00210 & 0.00500  &        &       &       & 7 &  $F_2$ & 5 & 6 & $F_1$ & 2 \\
 114.976348 &  1.0100 & 120/200/298 & 0.07341    &  0.02400  &  0.02442 &  0.700 &        &      &  0.07879  &   0.00300 & 0.00568  &        &       &       & 7 &  $E$  & 4 & 6 & $E$  & 2 \\
 124.762736 &  1.5340 & 200/298 & 0.05895    &  0.00740  &  0.00823 &  0.640 &        &      &  0.06571  &   0.00120 & 0.00420  &  0.866 & 0.150 & 0.159 &12 &  $E$  & 1 &11 & $E$  & 1 \\
 124.771153 &  2.3700 & 200/298 & 0.04521    &  0.00380  &  0.00470 &        &        &      &  0.06632  &   0.00091 & 0.00416  &  0.850 & 0.100 & 0.113 &12 &  $F_1$ & 1 &11 & $F_2$ & 2 \\
 124.783879 &  4.2670 & 200/298 & 0.03725    &  0.00170  &  0.00285 &  0.447 &  0.21  & 0.21 &  0.06296  &   0.00061 & 0.00390  &  0.867 & 0.062 & 0.082 &12 &  $A_1$ & 1 &11 & $A_2$ & 1 \\
 124.866856 &  2.6380 & 200/298 & 0.06200    &  0.00430  &  0.00574 &        &        &      & 
 0.07467  &   0.00091 & 0.00466  &  1.040 & 0.094 & 0.113 &12 &  $F_1$ & 2 &11 & $F_2$ & 3 \\
 124.881719 &  0.8196 & 200/298 & 0.03147    &  0.00670  &  0.00697 &  0.690 &        &      &  0.05150  &   0.00280 & 0.00422  &        &       &       & 9 &  $F_2$ & 3 & 8 & $F_1$ & 2 \\
 124.881895 &  0.8951 & 200/298 & 0.06800    &           &          &  0.690 &        &      &  0.11940  &   0.00750 & 0.01047  &        &       &       & 9 &  $F_1$ & 3 & 8 & $F_2$ & 2 \\
 124.909811 &  3.7530 & 200/298 & 0.04404    &  0.00220  &  0.00348 &  0.686 &  0.23  & 0.23 &  0.06528  &   0.00064 & 0.00405  &  0.994 & 0.069 & 0.092 &12 &  $F_2$ & 1 &11 & $F_1$ & 2 \\
 124.953539 &  2.2530 & 200/298 & 0.04981    &  0.00570  &  0.00647 &  3.090 &  0.38  & 0.42 &  0.06625  &   0.00098 & 0.00417  &  1.190 & 0.120 & 0.141 &12 &  $E$  & 2 &11 & $E$  & 2 \\
 124.958857 &  3.1670 & 200/298 & 0.04861    &  0.00360  &  0.00467 &        &        &      &  0.07090  &   0.00081 & 0.00442  &  1.020 & 0.083 & 0.104 &12 &  $F_2$ & 2 &11 & $F_1$ & 3 \\
 125.281451 &  1.0670 & 200/298 & 0.07404    &  0.01200  &  0.01283 &        &        &      &  0.07373  &   0.00180 & 0.00486  &  0.986 & 0.220 & 0.228 &12 &  $A_1$ & 2 &11 & $A_2$ & 1 \\
 134.958636 &  2.0730 & 200/298 & 0.04901    &  0.00350  &  0.00461 &  0.630 &        &      & 0.06542  &   0.00083 & 0.00409  &  0.733 & 0.120 & 0.128 &13 &  $F_2$ & 1 &12 & $F_1$ & 2 \\
 134.974968 &  2.1490 & 200/298 & 0.03682    &  0.00240  &  0.00329 &  0.630 &        &      & 0.06172  &   0.00076 & 0.00386  &  0.777 & 0.120 & 0.129 &13 &  $F_1$ & 1 &12 & $F_2$ & 1 \\
 135.064743 &  3.8510 & 200/298 & 0.05742    &  0.00270  &  0.00443 &  1.750 &  0.23  & 0.25 & 0.07541  &   0.00063 & 0.00466  &  0.814 & 0.078 & 0.093 &13 &  $A_1$ & 1 &12 & $A_2$ & 1 \\
 135.128296 &  2.7510 & 200/298 & 0.04496    &  0.00310  &  0.00415 &  1.870 &  0.33  & 0.35 & 0.06834  &   0.00074 & 0.00425  &  1.040 & 0.110 & 0.127 &13 &  $F_1$ & 2 &12 & $F_2$ & 2 \\
 135.136144 &  2.2070 & 200/298 & 0.03666    &  0.00280  &  0.00359 &  0.630 &        &      & 0.06317  &   0.00080 & 0.00395  &  1.040 & 0.120 & 0.136 &13 &  $E$  & 1 &12 & $E$  & 2 \\
 135.188872 &  2.5250 & 200/298 & 0.04372    &  0.00300  &  0.00402 &  1.090 &  0.37  & 0.38 & 0.07436  &   0.00083 & 0.00463  &  0.773 & 0.110 & 0.120 &13 &  $F_1$ & 3 &12 & $F_2$ & 3 \\
 135.205313 &  2.7630 & 200/298 & 0.04107    &  0.00250  &  0.00355 &  0.630 &        &      & 0.06756  &   0.00085 & 0.00422  &  1.080 & 0.110 & 0.128 &13 &  $F_2$ & 2 &12 & $F_1$ & 3 \\
 135.241444 &  5.2210 & 200/298 & 0.04882    &  0.00160  &  0.00339 &  0.987 &  0.19  & 0.20 & 0.07455  &   0.00054 & 0.00460  &  0.984 & 0.061 & 0.086 &13 &  $A_2$ & 1 &12 & $A_1$ & 2 \\
 135.296672 &  0.7780 & 200/298 & 0.04080    &  0.00750  &  0.00791 &  0.630 &        &      & 0.06380  &   0.00180 & 0.00430  &  0.974 & 0.310 & 0.316 &13 &  $F_1$ & 3 &12 & $F_2$ & 2 \\
 \hline\hline
\end{tabular}}

\end{table*}
\end{landscape}

\section{Rotational dependence of the measured parameters}\label{sec:rot_dep}

The measured self and N$_2$ broadening coefficients and their temperature dependence exponents are plotted in Fig.~\ref{fig:sel_n2_br} as a function of $|m|$ ($m=J''+1$ for \emph{R}-branch lines, $J''$ being the lower state rotational quantum number). Different markers and colors are used to distinguish the $A$-, $E$- and $F$-species transitions, showing no emerging trend with symmetry. All the measured quantities were therefore fitted to the following polynomial expansion
\begin{equation}\label{eq:polynome}
y = a + b \, |m|,
\end{equation}
each measured value being weighted according to the inverse of the square of its estimated overall uncertainty. The values of the fitted coefficients $a$ and $b$ obtained are provided in Table~\ref{Tab:coefficients}. Five outliers, $\gamma_{\mathrm{self}}$ = \SI{0.1194 \pm 0.0075}{\wn \atmosphere\tothe{-1}} at $|m|=9$, $\gamma_{\mathrm{N_2}}$ = \SI{0.38 \pm 0.15}{\wn \atmosphere\tothe{-1}} at $|m|=7$, $n_{\mathrm{self}}=\SI{1.83 \pm 0.21}{\nothing} $ at $|m|=7$, $n_{\mathrm{self}}=\SI{1.60 \pm 0.15}{\nothing} $ at $|m|=9$ and $n_{\mathrm{N_2}}$ = \SI{6.4 \pm 1.1}{\nothing} at $|m|=8$, are off scale. These data were excluded from the polynomial fits. The dispersion of the measured N$_2$ broadening is larger than that obtained for self broadening, most probably because the stability of the cell optics is more affected at the higher pressures used for N$_2$ broadening retrieval (minute shifts in the multipass optics occurred unavoidably after filling the cell with the warm gas mixtures). As a results shifts and channelling in the baseline increase the uncertainty in the line fitting. The same comment applies to N$_2$ temperature dependence exponents. Actually, even if a linear trend emerges, it is difficult to assert it with certainty, especially for $n_{\mathrm{N_2}}$.

\renewcommand{\arraystretch}{1.25}
\begin{table}[!ht]
\caption{Values of the fitted coefficients $a$ and $b$ obtained with the polynomial expression $y = a + b \, |m|$.}\label{Tab:coefficients}
\vspace{5pt}
\centering
\resizebox{0.8\columnwidth}{!}{\begin{tabular}{
	l
	S[table-format=1.3(3)]
	S[table-format=-1.4(3)]
	}
\hline\hline
 \multicolumn{1}{c}{$y$} & \multicolumn{1}{c}{$a^\text{a}$} & \multicolumn{1}{c}{$b^\text{a}$} \\
 \hline
$\gamma_\mathrm{self} / \si{\wn \atmosphere\tothe{-1}}$ & 0.088 \pm 0.009 & -0.0016 \pm 0.0008 \\
$\gamma_{\mathrm{N_2}} / \si{\wn \atmosphere\tothe{-1}}$ & 0.045 \pm 0.002 & \multicolumn{1}{c}{$0^\text{b}$} \\
$n_{\mathrm{self}} / \mathrm{unitless}$ & 0.641 \pm 0.211 & 0.0202 \pm 0.0196\\
$n_{\mathrm{N}_2} / \mathrm{unitless}$ & 0.506 \pm 0.840 & 0.0554\pm 0.0760 \\
\hline\hline
\multicolumn{3}{l}{\footnotesize{$^\text{a}$ Estimated uncertainty in parentheses.}} \\
\multicolumn{3}{l}{\footnotesize{$^\text{b}$ Parameter not fitted.}}
\end{tabular}}
\end{table}

\section{Discussion}\label{sec:discussion}

To the best of our knowledge, the present work involves the first measurements of N$_2$ broadening coefficients and their temperature dependence exponents for pure rotational lines of $^{12}$CH$_4$ at low temperature. So far, there are values of $\gamma_{\mathrm{air}}$ reported in HITRAN that come from references~\cite{brown2003methane, brown2005empirical, sanzharov2012self}. In addition, measurements were performed in the $\nu_4$ band by \citet{smith2009multispectrum} for air-broadened half widths, pressure-induced shifts and temperature dependence over a wide range of rotational levels. Similarly, the values used in HITRAN for $n_{\text{N}_2}$ are estimated temperature exponents described in Ref.~\cite{brown2003methane} and Ref.~\cite{brown2005empirical}.

\begin{figure}[ht]
  \centerline{\resizebox{1.0\hsize}{!}{\includegraphics{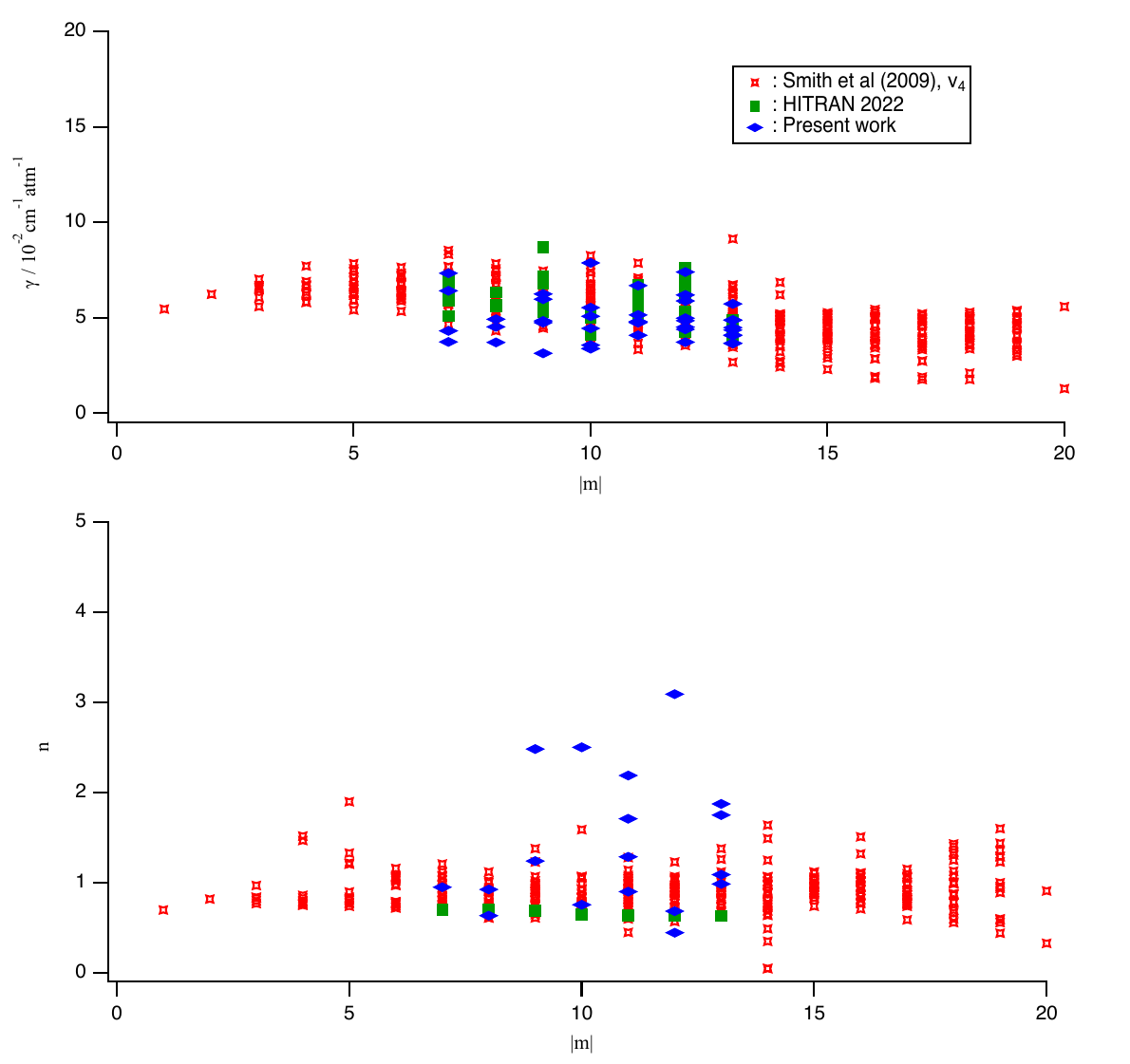}}}
\caption{Comparison of our measurements with the ones performed by \citet{smith2009multispectrum} of air-width and temperature dependence exponents in the $\nu_4$ band and the data found in HITRAN, in the [0, 20] range of $|m|$. One data point of this present work was removed in each graph for better visualization: at $m=|7|$ for the comparison of broadening coefficients, and at $m=|8|$ for the temperature dependence.}
\label{fig:Smith_comparison}
\end{figure}

Our results can thus be compared to the previous ones for $\nu_4$ and to those used in HITRAN, as performed in Fig.~\ref{fig:Smith_comparison}. While our data appear to suffer from significant deviation due to the difficulty of achieving the stability of the cell optics, and even though we had to remove one (obviously outlier, see Fig.~\ref{fig:sel_n2_br}) datum point in each graph for better visualization, the comparison shows a good agreement between air-width parameters and our N$_2$ parameters, even though with regard to the temperature dependence coefficients, it is more difficult to judge the quality of the comparison. Finally, as already outlined, we find no correlation between the values of the coefficients and the type of symmetry of the lines as illustrated in Fig.~\ref{fig:sel_n2_br}.

\section{Application to analysis of Titan’s spectra}\label{sec:titan}
We applied the measured broadening coefficients to an analysis of CH$_4$ rotational lines observed on Titan by the Cassini Composite InfraRed Spectrometer (CIRS) in the 70--\SI{150}{\wn} wavenumber range, aiming at deriving the stratospheric CH$_4$ volume mixing ratio (VMR) following the pioneering investigation of \citet{lellouch2014distribution}. We used a selection of Focal Plane 1 (FP1) spectra recorded on February 1st 2016 during the T116 flyby, with the field of view (FOV) centered in a latitude range of \SI{0.5}{\degree}S -- \SI{0.5}{\degree}N. The selection consists of 204 spectra having a mean latitude of \SI{0.4}{\degree}N, a FOV of about \SI{20}{\degree} in latitude, and a mean emission angle of \SI{43}{\degree}. The spectral resolution is \SI{0.52}{\wn}.

\begin{figure}[ht]
  \centerline{\resizebox{1.0\hsize}{!}{\includegraphics{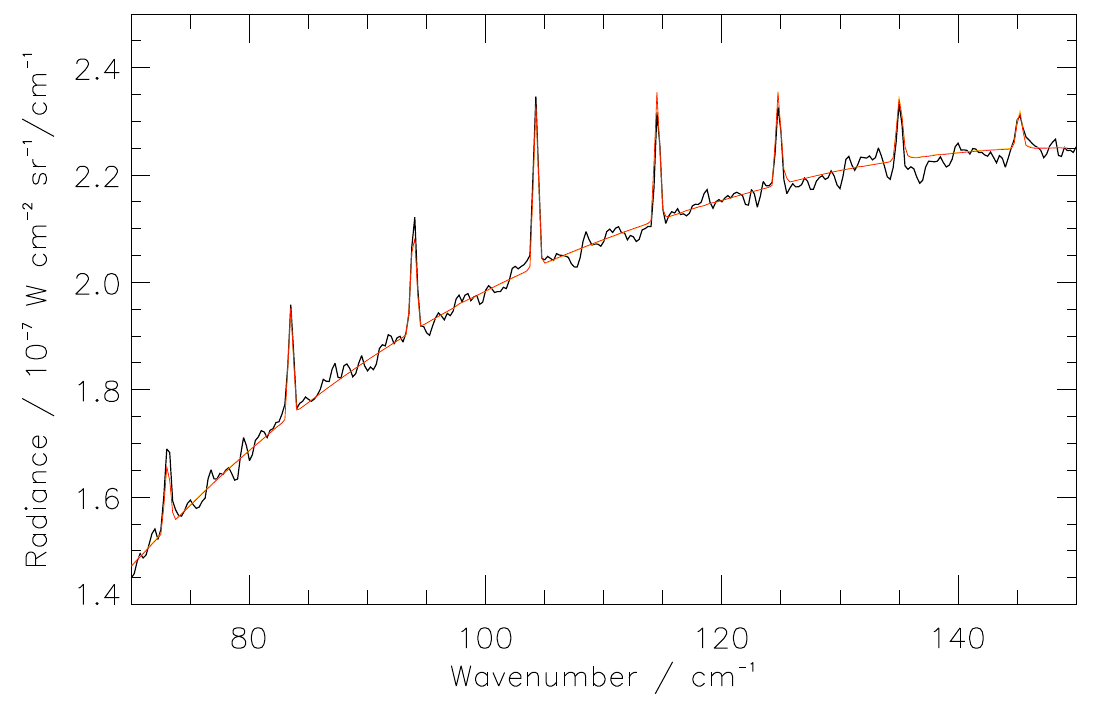}}}
\caption{Comparison of a selection of Cassini/CIRS spectra recorded near the equator in February 2016 (black) with a synthetic spectrum calculated with a CH$_4$ mole fraction of 1.17\% in the stratosphere. The temperature profile used in the calculations was derived by fitting the continuum between the CH$_4$ lines and the methane band observed by Cassini/CIRS around \SI{1305}{\wn}. The 1-standard deviation noise level in the observed spectrum is about \SI{7E-10}{\watt \centi \meter \tothe{-2} \steradian \tothe{-1} / \wn} and the spectral resolution is \SI{0.52}{\wn}.}
\label{fig:titan}
\end{figure}

Several methane lines are clearly seen over a continuum arising from the N$_2$-N$_2$ and CH$_4$-N$_2$ collision-induced absorption (and to a lesser extent from that of CH$_4$-CH$_4$ and N$_2$-H$_2$) and from the photochemical haze (Fig.~\ref{fig:titan}). The intensity of these rotational lines depends on both stratospheric CH$_4$ VMR and temperature profiles. In addition, it is thus necessary to use the $\nu_4$ band of methane at \SI{1305}{\wn}, which is more sensitive to temperature than to methane abundance. To do so, we made a selection of 453 Focal Plane 4 (FP4) CIRS spectra recorded between latitudes of \SI{5}{\degree}S and \SI{5}{\degree}N during the same flyby as the FP1 spectra. Its main emission angle is \SI{32}{\degree}. We then applied an iterative process to retrieve simultaneously the CH$_4$ VMR and the temperature profile from the fit of the 70--\SI{150}{\wn} range and of the $\nu_4$ band spectrum. The inversion algorithm that incorporates a line-by-line radiative transfer code is described in \citet{vinatier2015seasonal}. Opacity from the photochemical haze was calculated from the haze extinction profile derived by \citet{vinatier2020temperature} from limb spectra at \SI{5}{\degree}N acquired on February 1st 2016 with the spectral dependence of \citet{vinatier2012optical}. The CH$_4$ line positions, intensities and energy levels come from HITRAN2020 \citep{gordon2022hitran2020} and are similar to those used by \citet{lellouch2014distribution}. The N$_2$-N$_2$ collision-induced absorption was taken from \citet{gordon2022hitran2020}, as calculated from the trajectory-based approach of \citet{chistikov2019simulation}, and that of CH$_4$-N$_2$ comes from the new semi-empirical model presented in \citet{finenko2022trajectory}. For the room-temperature N$_2$ broadening coefficients and temperature exponent of the CH$_4$ lines, we used the $|m|$ dependence given by the polynomial fits in Table~\ref{Tab:coefficients}.

In a first step, we determined the temperature profile assuming an initial CH$_4$ VMR equal to 1.48\%, as measured by the Huygens GCMS \citep{niemann2010composition}, and taking the temperature profile of \citet{vinatier2020temperature} derived from CIRS spectra recorded near \SI{0}{\degree}N on February 1st 2016 as an \emph{a priori} in the inversion process. We fitted the $\nu_4$ CH$_4$ band centered at \SI{1305}{\wn} to infer stratospheric temperatures and the 70--\SI{280}{\wn} range to infer tropospheric ones (from the collision-induced continuum excluding the CH$_4$ rotational lines in this step). We then obtained a new temperature profile that was used, in a second step, to fit the CH$_4$ rotational lines in the 70--\SI{150}{\wn} spectral range. The derived CH$_4$ mixing ratio was then used as an input of a new iteration to retrieve the temperature profile in a first step and the CH$_4$ VMR in a second step. After a few iterations, we obtained convergence of both temperature and CH$_4$ abundance.

Fig.~\ref{fig:titan} shows the fit of the CH$_4$ rotational lines using the line broadening measurements presented here. The retrieved CH$_4$ VMR is (1.17 $\pm$ 0.08)\%. This value pertains to a region in the lower stratosphere centered at \SI{85}{\kilo \meter} (\SI{15}{\milli \bar}) \citep{lellouch2014distribution}. The inferred VMR is very close to the one obtained when the air-broadening coefficients from HITRAN2020 (based on values from other vibrational CH$_4$ bands) are used instead: 1.18\% with similar error bars. We can also note that the VMR we inferred agrees within error bars with that derived at the equator by \citet{lellouch2014distribution} from Cassini/CIRS nadir spectra recorded on December 5th 2008, (1.00 $\pm$ 0.10)\%, more than 7 years before the data analyzed here.

\section{Conclusion}\label{sec:conclusion}
Multi-spectrum analyses of pure rotational transitions of methane have been performed for pure and nitrogen diluted gas mixing at low temperature. These are the first results of this type and may be of great interest for the study of the radiative transfer of atmospheres such as Titan. The N$_2$ broadening parameters are in agreement with air broadening coefficients, obtained from other bands, reported in previous studies. This is less true for the temperature dependence exponents for which higher discrepancies are observed. Nevertheless, these measurements have been applied to observed spectra on Titan by the Cassini Composite InfraRed Spectrometer. Although the results do not show a big difference between these calculations and those made with the values already present in HITRAN, it is a valuable addition to the methane databases.

The measurement of hydrogen broadening parameters is a future experimental prospect in order to better characterize atmospheres that are mainly composed of hydrogen such as the giant planets Neptune and especially Uranus which should be visited by a space probe in a not too distant future.

Finally, these results should be integrated  very soon into the database of the Dijon group~\cite{RBR20}, which can be downloaded at \url{https://vamdc.icb.cnrs.fr/PHP/methane.php} and is also available through the VAMDC infrastructure~\cite{dubernet2010virtual,dubernet2016virtual,moreau2018vamdc,atoms8040076}.

\section*{Acknowledgments}
We acknowledge support from Synchrotron SOLEIL (projects 99180032 and 2019 0174) which enabled us to carry out these experiments. We acknowledge support from the French "Programme National de Planétologie" of INSU/CNRS.
%

\bibliographystyle{unsrtnat}
 \bibliography{biblio}


\end{document}